\documentclass[sigconf]{acmart}

\usepackage{multirow}
\usepackage{booktabs}
\usepackage{amsfonts} 
\usepackage{bm}
\usepackage{subcaption}
\usepackage{balance}

\newcommand{\bA}{\mathbf{A}}

\newcommand{\bD}{\mathbf{D}}
\newcommand{\bE}{\mathbf{E}}

\newcommand{\bH}{\mathbf{H}}

\newcommand{\bK}{\mathbf{K}}

\newcommand{\bM}{\mathbf{M}}

\newcommand{\bQ}{\mathbf{Q}}

\newcommand{\bS}{\mathbf{S}}

\newcommand{\bU}{\mathbf{U}}
\newcommand{\bV}{\mathbf{V}}
\newcommand{\bW}{\mathbf{W}}

\newcommand{\bb}{\mathbf{b}}

\newcommand{\be}{\mathbf{e}}

\newcommand{\bs}{\mathbf{s}}

\newcommand{\bu}{\mathbf{u}}

\newcommand{\bx}{\mathbf{x}}

\newcommand{\ie}{\textit{i.e.}}
\newcommand{\eg}{\textit{e.g.}}
\newcommand{\norm}[1]{\left\lVert#1\right\rVert}

\setcopyright{acmcopyright}

\copyrightyear{2021} 
\acmYear{2021} 
\setcopyright{acmlicensed}\acmConference[CIKM '21]{Proceedings of the 30th ACM International Conference on Information and Knowledge Management}{November 1--5, 2021}{Virtual Event, QLD, Australia}
\acmBooktitle{Proceedings of the 30th ACM International Conference on Information and Knowledge Management (CIKM '21), November 1--5, 2021, Virtual Event, QLD, Australia}
\acmPrice{15.00}
\acmDOI{10.1145/3459637.3482269}
\acmISBN{978-1-4503-8446-9/21/11}

\acmSubmissionID{68}

\begin{document}
\fancyhead{}
\title{Learning Implicit User Profiles for Personalized Retrieval-Based Chatbot}


\author{Hongjin Qian\textsuperscript{\rm 1}, Zhicheng Dou\textsuperscript{\rm 1*}, Yutao Zhu\textsuperscript{\rm 3}, Yueyuan Ma\textsuperscript{\rm 1}, Ji-Rong Wen\textsuperscript{\rm 1,2}}
\affiliation{
\textsuperscript{\rm 1} Gaoling School of Artificial Intelligence, Renmin University of China \state{Beijing} \country{China} \\
\textsuperscript{\rm 2} Beijing Key Laboratory of Big Data Management and Analysis Methods \state{Beijing} \country{China} \\
\textsuperscript{\rm 3} Université de Montréal \state{Québec} \country{Canada}}
\email{{ian, *dou}@ruc.edu.cn}
\renewcommand{\authors}{Hongjin Qian, Zhicheng Dou, Yutao Zhu, Yueyuan Ma, and Ji-Rong Wen}
\renewcommand{\shortauthors}{Qian, et al.}
\begin{abstract}
In this paper, we explore the problem of developing personalized chatbots. A personalized chatbot is designed as a digital chatting assistant for a user. The key characteristic of a personalized chatbot is that it should have a consistent personality with the corresponding user. It can talk the same way as the user when it is delegated to respond to others' messages. Many methods have been proposed to assign a personality to dialogue chatbots, but most of them utilize explicit user profiles, including several persona descriptions or key-value-based personal information. In a practical scenario, however, users might be reluctant to write detailed persona descriptions, and obtaining a large number of explicit user profiles requires tremendous manual labour. To tackle the problem, we present a retrieval-based personalized chatbot model, namely IMPChat, to learn an implicit user profile from the user's dialogue history. We argue that the implicit user profile is superior to the explicit user profile regarding accessibility and flexibility. 
IMPChat aims to learn an implicit user profile through modeling user's \textbf{personalized language style} and \textbf{personalized preferences} separately. To learn a user's personalized language style, we elaborately build language models from shallow to deep using the user's historical responses; To model a user's personalized preferences, we explore the conditional relations underneath each post-response pair of the user. The personalized preferences are dynamic and context-aware: we assign higher weights to those historical pairs that are topically related to the current query when aggregating the personalized preferences. 
We match each response candidate with the personalized language style and personalized preference, respectively, and fuse the two matching signals to determine the final ranking score. We conduct comprehensive experiments on two large datasets, and the results show that our method outperforms all baseline models. The codes and datasets will be released at Github\footnote{https://github.com/qhjqhj00/CIKM2021-IMPChat}.


\end{abstract}




\maketitle

\section{Introduction}\label{sec:intro}
Building an open-domain dialogue system is an intriguing but challenging task that draws more and more attention in both academic and industrial communities. In general, there are two pathways to construct an open-domain chatbot: retrieval-based and generation-based. The former usually retrieves relevant response candidates via a retrieval engine and then selects a proper response from these candidates~\cite{wang-etal-2013-dataset, Wu2017, zhou-etal-2018-multi, tao-etal-2019-one, Lowe2015}. The latter directly leverages models, such as encoder-decoder, to generate response texts~\cite{wang-etal-2013-dataset, DBLP:journals/corr/VinyalsL15, DBLP:conf/emnlp/RitterCD11}. In this paper, we concentrate on the retrieval-based chatbots which are wildly applied in many industrial products such as Microsoft XiaoIce~\cite{zhou2020design} and E-commerce assistant AliMe~\cite{DBLP:conf/cikm/LiQCWGHRZZWJC17}.

Although plentiful chatbot applications are available, there still exist some challenges that cannot be overlooked~\cite{Chaves_2020,gao2019neural}. Inconsistent personality is one of the most mentioned challenges of current open-domain chatbots~\cite{shum2018eliza, de2001unfriendly}. In the dialogue chatbot domain, personality refers to a set of traits that define a chatbot's interaction style, character, and behaviors~\cite{de2001unfriendly, liu-etal-2020-speaker}. Inconsistent personality brings more unpredictability and untrustworthiness to the chatbot, which can disorient end users and lead to a strong sense of discomfort~\cite{de2001unfriendly, 10.1145/3123818.3123826}. Therefore, endowing chatbots with consistent personality becomes a crucial research task that guarantees a chatbot's performance in agreement with user's expectations. To this end, many methods have been proposed to assign a personality to dialogue chatbots~\cite{DBLP:journals/corr/Al-RfouPSSSK16,DBLP:conf/emnlp/BakO19,DBLP:conf/acl/LiGBSGD16,DBLP:conf/acl/KielaWZDUS18,qian2017assigning}. In this paper, we address the problem of developing ``personalized chatbots''.  A personalized chatbot is created for a specific user and it functions like the digital chatting assistant of the user. Ideally, it can talk exactly the same way as the user, and can respond to others' messages when it is delegated by the user. We argue that for such a personalized chatbot, it is extremely important to have a consistent personality with the corresponding user. 

There have been a few attempts for developing personalized chatbots. Early studies encoded user information by using user ID embeddings which are learned during training~\cite{DBLP:journals/corr/Al-RfouPSSSK16,DBLP:conf/emnlp/ChanLYCHZY19,DBLP:conf/acl/LiGBSGD16}, and generate personalized responses with the help of the ID embeddings. Recently, some works focused on building personalized chatbots using manually created user profiles to maintain the consistency of personality for chatbots~\cite{DBLP:conf/acl/LiuCCLCZZ20,mazare-etal-2018-training,DBLP:conf/acl/SongWZLL20,DBLP:conf/aaai/SongZH020,Zhang2018,gu2019dually,10.1145/3340531.3411967}. Such predefined user profiles are \textit{explicit} user profiles which usually contain several persona descriptions or key-value-based personal information. 

\begin{table}[t!]
\small
\centering
\caption{An example of a user's dialogue history.}
\begin{tabular}{p{.65\linewidth}|p{.25\linewidth}} 
\toprule
Post & Response \\ 
\midrule
  (1) What's your favorite programming language? & (1) Mine is Java.   \\
  (2) Rafa wins his 1000 match! & (2) Bravo Rafa!    \\
  (3) PC or MAC for college students? & (3) If ur in steam, go windows.    \\
  (4) I failed an exam again and feel like a loser. & (4) Good for you.    \\
  (5) Australian Open Final Nadal vs Federer! & (5) Vamos, Nadal.    \\
\bottomrule
\end{tabular}
\label{tab:user}
\end{table}

In this paper, we propose using the user's dialogue history to automatically build \textbf{implicit user profiles} to endow consistent personality to the chatbot. The reasons mainly lie in two aspects.
\textbf{First}, regarding accessibility and flexibility, the implicit user profile is superior to the explicit user profile. Obtaining a large number of explicit user profiles requires tremendous manual labour~\cite{DBLP:conf/emnlp/ChanLYCHZY19}. In a practical scenario, users also might be reluctant to write such detailed persona descriptions~\cite{DBLP:journals/misq/GerlachK91}. Besides, compared with explicit user profiles, implicit user profiles are easier to update with the accumulation of dialogue history as they are automatically learned.
\textbf{Second}, tremendous personalized style information is usually hidden underneath the user's dialogue history. Here, the personalized style refers to the unique personal characteristics (\eg, speaking style, background knowledge, and preferences) that help distinguish a user from others. Such personalized style information is beneficial to enhance the chatbot's ability to output coherent responses. Table~\ref{tab:user} shows an example of a user's dialogue history from which we can find the user's personalized speaking style and preferences.
Taking the fifth post ``Australian Open Final Nadal vs Federer!'' as an example, the user replies ``Vamos\footnote{``Vamos'' is a Spanish word which means ``let's go''. }, Nadal''. Thus, we know that the user likes the tennis player Nadal instead of Federer. And the user likes using ``Vamos'' to cheer the sports player on. 



Inspired by the beneficial properties of implicit user profiles, we design a model {IMPChat} to learn the \textbf{IM}plicit user profile from a user's dialogue history for \textbf{P}ersonalized retrieval-based \textbf{Chat}bots. Through the model, we propose modeling the implicit user profile from two aspects: personalized language style and personalized preference. The former aims to summarize personalized language characteristics (\eg, using ``Vamos'' to cheer sports player on) from the dialogue history without considering the current message. The latter considers the conditional relation between the current message and the user's personal preference (\eg, like the tennis player Nadal instead of Federer) by building a post-aware user profile. Then, we perform matching between the response candidates and the learned implicit user profile to select a personalized response.


Specifically, we \textbf{first} design a \textbf{Personalized Style Matching} module that uses multi-layer attentive modules to capture multi-grained personalized language characteristics. At each layer, we simultaneously perform self-attention within historical responses and cross-attention with the corresponding historical posts. Through this way, we not only obtain deep personalized style features but understand the context-aware relations between posts and responses. We perform matching between the response candidates and the personalized style.
\textbf{Second}, we build a \textbf{Post-Aware Personalized Preference Matching} module. Given the current query, this module first selects topically related dialogue history. Then, it models the personalized preferences from the selected dialogue history to enhance the personalized response selection. The dialogue history selection is performed by measuring the topical relatedness between the current query and each historical post. We think that the word ambiguity of the current query might bias the topical relatedness. To deal with the bias, we create a tailored multi-hop method to stretch the current query's semantic richness by fusing context information into the current query. We then perform matching between the response candidates and the post-aware personalized preference.
\textbf{Finally}, in the fusion module, we combine the two 
matching features to determine the most proper response. 
To validate the effectiveness of our model, we conduct comprehensive experiments on two publicly available datasets for personalized response selection. Experimental results show that our model outperforms all baseline models.

Our contributions are threefold: 
(1) To the best of our knowledge, this is the first study that seeks to build an implicit user profile using the user's dialogue history to enhance personality consistency for the retreival-based chatbot. 
(2) We propose {IMPChat} that constructs an implicit user profile considering the user's personalized language style and personalized preference. 
(3) Extensive experiments show that our model outperforms the state-of-the-art models.

\section{Related Work}
\subsection{Retrieval-Based Chatbots}
To build an open-domain dialogue chatbot, there exist two major research directions: generation-based and retrieval-based. The first usually learns to generate responses with an encoder-decoder structure~\cite{10.3115/1117562.1117568,DBLP:journals/corr/VinyalsL15,DBLP:conf/acl/ShangLL15,DBLP:conf/naacl/LiGBGD16,DBLP:conf/aaai/SerbanSBCP16,DBLP:journals/ir/ZhuDNW20}. The second learns a matching model to select proper responses from candidates given input query and context~\cite{Lowe2015,zhou-etal-2016-multi,DBLP:conf/acl/WuWXZL17,tao-etal-2019-one,DBLP:conf/acl/ZhuSDNZ20,DBLP:conf/sigir/ZhuNZDJD21}. In this paper, we focus on the 
latter.

The general objective of retrieval-based methods is to select a proper response from a candidate list based on a given query\footnote{Here, ``query'' has different meanings with that often used in IR.}. The candidates are usually pre-retrieved by an index system. 
Along this direction, early studies mainly focus on single-turn response selection~\cite{hu2014convolutional,wang-etal-2013-dataset,wang2015syntax,10.1145/3159652.3159659}. Recently, study interests move to multi-turn response selection, which uses dialogue session information as context to enhance the single-turn matching. For example, Dual-LSTM model encodes context and responses into hidden states and the last hidden states are used to compute relevance score~\cite{Lowe2015}. Multi-view model~\cite{zhou-etal-2016-multi} performs multi-grained context-response matching. These methods concatenate the context to a long sequence which may undermine the relationships among utterances. To tackle this problem, the representation-matching-aggregation framework is then applied by many recent works, which conducts interaction between the response and each utterance in the context and then aggregate these matching features using CNN or RNN~\cite{DBLP:conf/acl/WuWXZL17}. Following the framework, deep utterance aggregation network (DUA)~\cite{zhang2018modeling} designs turns-aware aggregation mechanism that assigns different weights to each dialogue turn. Deep attentive matching network (DAM) jointly uses self-attention and cross-attention to obtain multi-grained representations~\cite{zhou-etal-2018-multi}. Multi-hop selection network (MSN)~\cite{yuan-etal-2019-multi} designs a multi-hop method to make the context selection more robust. Besides, the interaction-over-interaction network (IoI) argues that one time of interaction captures limited matching features, and the IoI network stacks multiple interaction blocks~\cite{tao-etal-2019-one}. 

\subsection{Personalized Chatbots}
Endowing personality to open-domain chatbots is intriguing but challenging. It is an inevitable barrier to achieve truly applicable intelligent assistant. With a consistent personality, the ability of a chatbot to be precise regarding how to use language would be greatly improved~\cite{morrissey2013realness,DBLP:conf/sigir/MaDZZW21}. Early work tries to model personality using heuristic methods such as ``Big Five'' of speakers~\cite{DBLP:conf/acl/MairesseW07} . With the rising of deep learning recently, many end-to-end neural models are designed to tackle the challenge. Basically, these methods can be categorized into two groups. The first group learns a user embedding for each user. The user embedding is then used to conduct matching or generation~\cite{DBLP:journals/corr/Al-RfouPSSSK16,DBLP:conf/emnlp/BakO19,DBLP:conf/acl/LiGBSGD16}. Albeit the user embeddings are updating through training, they cannot directly interact with dialogue texts which contain abundant personalized information. 

The second group learns personality using explicit user profile (persona) which contains either several persona sentences or key-value agent profile~\cite{DBLP:conf/acl/KielaWZDUS18,qian2017assigning}. These explicit user profiles are either human-annotated or extracted from the dialogue history using rule-based methods. PERSONA-CHAT is such a dataset that is widely used~\cite{DBLP:conf/acl/KielaWZDUS18}. The major problem is how to use persona sentences (documents) and the context properly. Along with the dataset, a key-value profile memory network is used, which takes the dialogue history as input and performs attention over persona sentences~\cite{DBLP:conf/emnlp/MillerFDKBW16,DBLP:conf/acl/KielaWZDUS18}. DGMN lets the context and document attend to each other to obtain fused representations~\cite{DBLP:conf/ijcai/ZhaoTWX0Y19}. CSN performs document selection over the persona sentences~\cite{ZhuNZDD21}. RSM-DCK conducts selection over both dialogue context and user profiles and lets the response candidates interact with the selected results~\cite{10.1145/3340531.3411967}. 

Constructing a dataset like PERSONA-CHAT is exhausted. Some other works tackle this issue by generating user profiles through rule-based methods~\cite{zhong2020towards,mazare-etal-2018-training}. For example, Mazaré et al.~\cite{mazare-etal-2018-training} construct much larger datasets with explicit user profile using Reddit\footnote{https://www.reddit.com/r/datasets/ comments/3bxlg7/} dataset in a heuristic way, and it empirically proves that models training on such automatically-generated user profile with enough data can achieve similar or even better performance. Besides, the topic of personalization has also been explored in many other scenarios such as personalized search and user modelling~\cite{DBLP:conf/sigir/ZhouDWXW21, DBLP:conf/wsdm/ZhouDW20, DBLP:conf/sigir/ZhouDW20}. 

Unlike the studies above, in this paper, we seek to build an implicit user profile using the user's dialogue history. Such implicit user profile contains rich personalized information and can be automatically learned from the historical dialogues. 


\section{Methodology}

Assigning personality to chatbots is crucial to satisfy user's expectations.  As introduced in Section~\ref{sec:intro}, most current works seek to model personality with explicit user profiles. In this paper, we instead study the problem of automatically modeling implicit profiles of a user based on the user's dialogue history, and leverage the implicit profile to select the personalized response. The chatbot acts like a virtual agent of the user and is able to reply messages when the user is busy and asks the chatbot to generate responses.
\subsection{Preliminary}
\label{sec:pro}
The general goal of a retrieval-based chatbot is to return the most proper response within a database for an input message. Assume $g$ is a scoring model evaluating the quality of a candidate response $r$ for a given message $q$ under the context $\mathcal{C}$, the chatbot will output the response $r^*$ with the highest value of $g$ from a repository of response $\mathcal{R}$, \ie, we have:
\[
r^* = \mathop{\arg\max}_{r\in \mathcal{R}} g(q, r, C). 
\]
For a single-turn chatbot, $\mathcal{C}=\phi$ and the response is selected merely based on the input message $q$. For a multi-turn chatbot, $C$ is the conversation context comprised of previous turns of dialogues. In this work, we investigate the problem of retrieving a personalized response that matches the speaking style and knowledge background of a user $u$. Suppose $\mathcal{U}$ is the corresponding profile of $u$, and we have $C=\mathcal{U}$. $\mathcal{U}$ is usually a compact representation of a user's interests and knowledge. Hence, the personalized chatbot aims to return the response with the highest value of $g(q, r, \mathcal{U})$. 

There are multiple ways to model $\mathcal{U}$. Previous works either learn user ID embeddings~\cite{DBLP:journals/corr/Al-RfouPSSSK16,DBLP:conf/emnlp/BakO19,DBLP:conf/acl/LiGBSGD16} or build explicit user profiles\cite{DBLP:conf/acl/KielaWZDUS18,DBLP:conf/ijcai/ZhaoTWX0Y19,ZhuNZDD21,10.1145/3340531.3411967}. For example, Zhang et al.~\cite{DBLP:conf/acl/KielaWZDUS18} used explicit persona descriptions and Qian et al.~\cite{qian2017assigning} used key-value profile to model $\mathcal{U}$. In this paper, we instead model $\mathcal{U}$ using the user's dialogue history. Suppose $H=\left\{\left(p_j,r_j\right)
\right\}, j\in[1,t]$ is the user's dialogue history where $p_j$ represents posts issued by others and $r_j$ means responses made by the current user, we propose selecting responses by:
\[
g(q, r, \mathcal{U}) = g(q, r, \mathcal{F}(H)),
\]
where $\mathcal{F}(\cdot)$ defines a model that learns implicit user profile from dialogue history. Note that we further denote the user's historical posts and responses in $H$ as $P=\{p_j\}$ and $R=\{r_j\}$ respectively.

\begin{figure*}
    \centering
    \includegraphics[width=\linewidth]{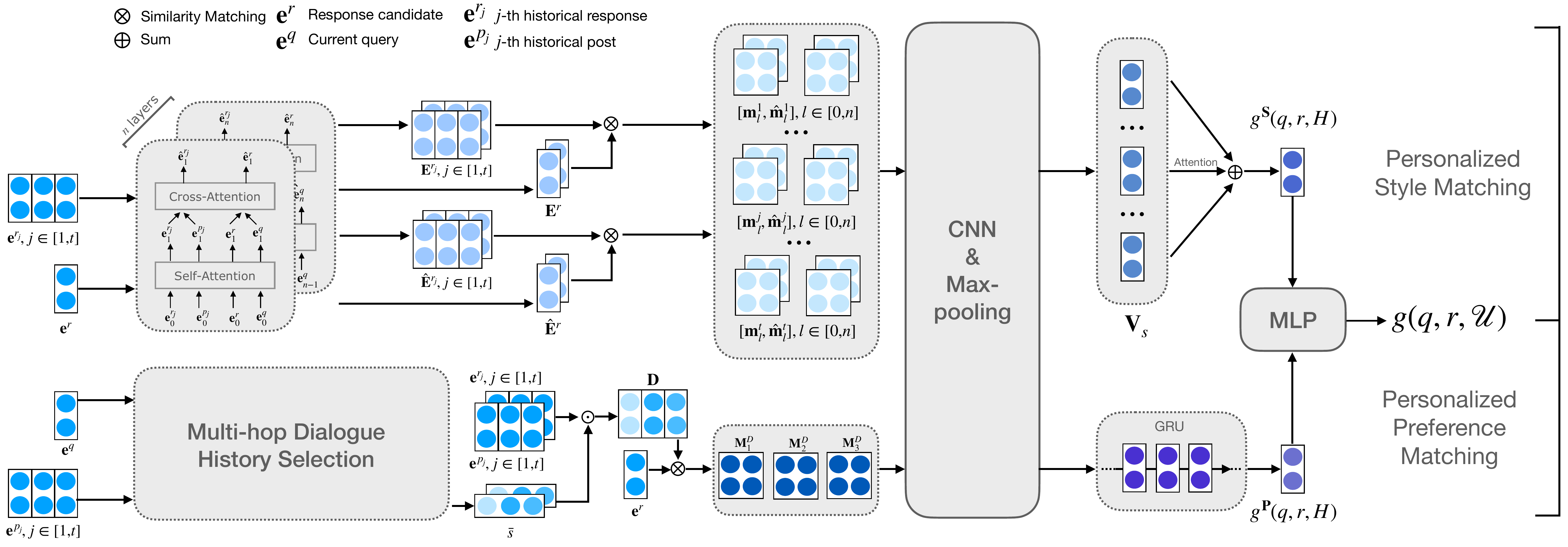}
    \caption{The overall Architecture of the proposed model IMPChat which is comprised of the Personalized Style Matching module and the Personalized Preference Matching module. The detail of Multi-hop module is at figure \ref{fig:hop}.}
    \label{model}
\end{figure*}
\subsection{IMPChat Overview}\label{themodel}

Intuitively, there are two major factors that impact the manner of expressing. First, a user has a personalized language style regardless of the conversation context. Such language style is usually determined by the user's knowledge background and preferred expressions. Second, the way a user makes responses is also conditioned on the intrinsic relations between a post and the user's personalized preferences. Given the same dialog context (e.g., a post), different users may give different responses and we try to learn the personalized preferences from history.

With this in mind, we assume the personalized user profile $\mathcal{U}$ is comprised of two parts: (1) the personalized language style \textbf{S} and (2) the post-aware personalized preferences \textbf{P}, and we have:
\[
g(q, r, \mathcal{U}) = m\left(g^{\textbf{S}}\left(q, r, H\right), g^{\textbf{P}}\left(q, r, H\right)\right),
\]
where $g^{\textbf{S}}$ and $g^{\textbf{P}}$ return the matching features of the candidate response $r$ regarding the two parts of the personalized user profile.

Figure~\ref{model} shows the structure of our proposed IMPChat. Specifically, we first build a \textbf{Personalized Style Matching} module (which will be introduced in Section \ref{sec:per}) which aims to capture the personalized language style using the user's historical responses and get $g^{\textbf{S}}$ which measures the style consistency of the response candidates. Then, we design a \textbf{Post-Aware Personalized Preference Matching} module (in Section \ref{pro}) which learns post-aware user preferences and get $g^{\textbf{P}}$ which measures the matching degree between the post-aware user preferences and the response candidates. The two modules perform matching separately, and in the \textbf{fusion module} (in Section~\ref{fus}), we combine the two matching features to compute the final matching score $g(q, r, \mathcal{U})$.
In the remaining parts of this section, we will introduce the details of each component.
 
 


\subsection{Foundation: Attentive Module}\label{sec:attentivemodule}
We first introduce the structure of Attentive Module, which is a basic component in our method.
Following the previous work~\cite{zhou-etal-2018-multi,yuan-etal-2019-multi}, we use the Attentive Module to endow contextual semantics into word embeddings. Attentive Module is proposed in~\cite{zhou-etal-2018-multi}. It is a variant of Transformer model~\cite{DBLP:conf/nips/VaswaniSPUJGKP17}. Instead of the multi-head attention that is used in the Transformer, the Attentive Module uses single-head attention. 
Specifically, the Attentive Module takes $\bQ\in\mathbb{R}^{t\times d}$, $\bK\in\mathbb{R}^{t\times d}$, and $\bV\in\mathbb{R}^{t\times d}$ as input, where $t$ and $d$ represent sequence length and embedding dimension respectively.

The Attentive Module defines an attention function ${\rm Att}(\bQ,\bK,\bV)$ to map the query $\bQ$ and key-value pair $(\bK,\bV)$ to a weighted output. The weights are computed by letting each word in the query sentence attend to words in the key sentence via scaled dot production~\cite{DBLP:conf/nips/VaswaniSPUJGKP17}, which can be formulated as:
\begin{equation}
    {\rm Att}(\bQ,\bK,\bV)={\rm softmax}(\frac{\bQ\cdot\bK^\top}{\sqrt{d}})\bV.
\end{equation}
A residual connection with layer normalization~\cite{ba2016layer} is then applied to get a better fused representation $\bx$ and prevent vanishing or exploding of gradients. A feed-forward network (FFN) with ReLU~\cite{goodfellow2016deep} activation is further applied to the normalized results as:
\begin{equation}
    {\rm FFN}(\bx) = {\rm ReLU}(\bx\cdot\bW_1+\bb_1)\cdot\bW_2+\bb_2,
\end{equation}
where $\bx$ is a 2D tensor with the same shape of the query $\bQ$; and $\bW_1$, $\bW_2$, $\bb_1$, and $\bb_2$ are parameters. The final output is obtained via a residual connection with normalization between ${\rm FFN}(\bx)$ and $\bx$. We denote the whole Attentive Module as $f_{\text{att}}(\cdot,\cdot,\cdot)$.


\subsection{Personalized Style Matching}
\label{sec:per}
A personalized chatbot should coherently output responses that portray consistent personalized styles (\eg, speaking style and vocabulary). We think that these personalized styles are very helpful when we expect to retrieve personalized responses. Taking the pet phrase (a type of speaking style) as an example, regarding how to express congratulation, some users might use ``neat'' or ``congrats'', while the user in Table~\ref{tab:user} would like to use ``bravo''. Users tend to use their preferred expressions more frequently, and these personalized preferred expressions are underneath the user's historical responses. Under such an observation, we design a \textit{Personalized Style Matching} module that aims to model a user's preferred speaking style from the user's historical responses. 


Formally, given a user $u$ with dialogue history $H=\left\{\left(p_j,r_j\right)
\right\}$ and the response candidate $r$, the Personalized Style Matching module aims to get a style matching feature vector $g^{\textbf{S}}\left(q, r, H\right)$ which measures the style consistency between the response candidate $r$ and the historical responses $R=\{r_j\},j\in[1,t]$.
The Personalized Style Matching module achieves the goal via three layers: (1) \textbf{Representation}: it extracts multi-grained semantic representations for historical responses and the response candidate; (2) \textbf{Matching}: it performs matching at each semantic level, and (3) \textbf{Aggregation}: it dynamically fuses matching signals between the response candidate and all historical responses to obtain $g^{\textbf{S}}\left(q, r, H\right)$. We will introduce each layer in detail as follows. 

The \textbf{representation layer} aims to obtain multi-grained contextual representations $\bE = \{\be_0,\cdots,\be_n\}$ and cross-attention representations $\hat{\bE} = \{\hat{\be}_0,\cdots,\hat{\be}_n\}$ for the response candidate $r$ and each historical response $r_j,j\in[1,t]$ 
using $n$ attentive modules.

Taking the $j$-th response $r_j$ as an example, the contextual representations $\bE^{r_j}=\{\be^{r_j}_0,\cdots,\be^{r_j}_n\}$ model the contextual semantic pattern of $r_j$ from shallow to deep. Specifically, we first initialize word representations $\be^{r_j}_0$ by looking up a word embedding table (\eg, Word2Vec). Then, we obtain deep contextual response representations by feeding the word embeddings into $n$ attentive modules:
\begin{equation}
\label{eq:att}
    \be^{r_j}_l =f_{\text{att}}(\be^{r_j}_{l-1}, \be^{r_j}_{l-1}, \be^{r_j}_{l-1}), \quad 1\leq l \leq n,
\end{equation}
where  $\be^{r_j}_l$ is the contextual representation output by the $l$-th attentive module.
Through $n$ attentive modules, we obtain the $n+1$ representations which depict the co-occurrence pattern of words at different granularities. 

Furthermore, we think that the personalized style of a response is also conditioned on the post. For example, ``good for you'' is semantically similar to ``bravo''. However, in Table~\ref{tab:user},  ``good for you'' is a response to the post ``I failed an exam again and feel like a loser''. The user instead uses the phrase in an ironical way. In view of this, we let the response $r_j$'s contextual representation $\be^{r_j}_l\in \bE^{r_j}$ attend to the corresponding post $p_j$'s contextual representation $\be^{p_j}_l\in \bE^{p_j}$ to obtain the cross-attention representations $\hat{\bE}^{r_j} = \{\hat{\be}^{r_j}_0,\cdots,\hat{\be}^{r_j}_n\}$: 
\begin{equation}
    \hat{\be}^{r_j}_{l} =f_{\text{att}}(\be^{r_j}_l, \be^{p_j}_l, \be^{p_j}_l), \quad l\in[0,n],
\end{equation} 
where $\bE^{p_j}$ is obtained in the same way as $\bE^{r_j}$. 
For the response candidate $r$, we obtain $\bE^{r}=\{\be^{r}_0,\cdots,\be^{r}_n\}$ and $\hat{\bE}^{r} = \{\hat{\be}^{r}_0,\cdots,\hat{\be}^{r}_n\}$ in the same way. 

The \textbf{matching layer} aims to obtain a personalized style matching matrix $\bM_s$ which measures the style matching degree between the response candidate and each historical response at multiple granularities. Specifically, given a candidate response $r$ and the $j$-th historical response $r_j$, we compute $\textbf{m}_l^j$ and $\hat{\textbf{m}}_l^j$ which are the matching matrices for the contextual representations $\{\be^{r_j}_l, \be^{r}_l\}$ and the cross-attention representations $\{\hat{\be}^{r_j}_l,\hat{\be}^{r}_l\}$:
\begin{align}
    \textbf{m}_l^j = \frac{\be^{r_j}_l \cdot {\be^{r\top}_l} }{\sqrt{d}}, \quad
    \hat{\textbf{m}}_l^j = \frac{\hat{\be}^{r_j}_l \cdot {\hat{\be}^{r\top}_l}}{\sqrt{d}}, \quad l\in [0,n],
\end{align}
where $d$ is the dimension of the embeddings and $l$ refers the representation output by the $l$-th attentive module. Hence, for $t$ historical responses $R=\{r_1,\cdots r_t\}$, we have two groups of multi-grained matching matrices $\bM_l=\{\textbf{m}_l^1,\cdots,\textbf{m}_l^t\}$ and $\hat{
\bM}_l=\{\hat{\textbf{m}}_l^1,\cdots,\hat{\textbf{m}}_l^t\}$.

To transform these matching matrices into a shared feature space, we first concatenate them into a stacked matching matrix:
\begin{align}
    \bM_s = f_{\text{stack}}\left(\{\bM_0,\cdots,\bM_n,\hat{\bM}_0,\cdots,\hat{\bM}_n\}\right),
\end{align}
where $f_{\text{stack}}(\cdot)$ refers to concatenation along a new dimension. $\bM_s \in \mathbb{R}^{2(n+1)\times t \times L \times L}$ and $L$ is the maximum sequence length.

Next, in the \textbf{aggregation layer}, following~\cite{zhou-etal-2018-multi,yuan-etal-2019-multi,ZhuNZDD21}, we extract matching features from the matching matrix $\bM_s$ via CNN. The extracted feature is then linearly mapped into a lower dimension:
\begin{align}
    \bV_s = {\rm MLP}\left(f_{\text{CNN}}\left({\bM_s}\right)\right),
\end{align}
where $\bV_s \in \mathbb{R}^{t\times d}$, and {\rm MLP}($\cdot$) represents a multi-layer perceptron.  

Note that the personalized style matching feature $\bV_s$ contains the matching signals between the response candidate $r$ and each historical response $r_j$. Although post-response pairs are sorted by time in the dialogue history, temporal patterns vanish for historical responses solely. Besides, each historical response may impact differently on personalized style. Therefore, we apply self-attention to dynamically sum up the personalized style matching feature $\bV_s$. Finally, we obtain the style matching features $g^{\textbf{S}}\left(q, r, H\right)$:
\begin{align}
   \bs_{\text{att}} &= {\rm softmax}({\rm MLP}(\tanh\left({\rm MLP}(\bV_s)\right))),\\
   g^{\textbf{S}}\left(q, r, H\right) &= \sum\limits_{dim=0}\bs_{\text{att}}\odot \bV_s,
\end{align}
where $\bs_{\text{att}}\in\mathbb{R}^{t\times d}$ represents the attention weights, and $\odot$ is the element-wise multiplication.

\subsection{Post-Aware Personalized Preference Matching}
\label{pro}

In addition to personalized language styles, a user's personalized preferences also have a great impact on the personalized response selection. For example, people who prefer MAC to PC would give a positive comment about a Mac-related post. 
We think that such personalized preferences tend to be captured from the user's dialogue history as the user's preferences are relatively consistent. 
Hence, how to properly utilize the dialogue history to enhance the personalized preference consistency is important.

The \textit{Post-Aware Personalized Preference Matching} module aims to obtain a matching vector $g^{\textbf{P}}\left(q, r, H\right)$ which measures whether a response candidate $r$ can consistently reflect the user's personalized preferences given the current query $q$ and a user $u$ with dialogue history $H$. To properly utilize the dialogue history $H$, we first transform the dialogue history $H$ into a user profile $\bD$ from which we can effectively model the personalized preferences. 


Ideally, the user profile $\bD$ only contains post-response pairs that are topically related to current post.  However, dialogue history reflects multifaceted user interests, some of which might be unrelated to current post. Unrelated context might bring negative impact on response selection~\cite{yuan-etal-2019-multi, 10.1145/3340531.3411967}. 
Therefore, we need to filter out the unrelated dialogue history. Intuitively, we can compute a relevance vector $\bs\in \mathbb{R}^t$ that measures the topical relatedness between the current post $q$ and historical posts $P=\{p_1,\cdots,p_t\}$. Then,we can obtain the user profile $\bD$ by reweighting the dialogue history $\bH=\left\{\left(\be^{p_j},\be^{r_j}\right)
\right\}, j\in[1,t]$:
\begin{equation}
\label{eq:reweight}
    \bD=\bs\cdot\bH,
\end{equation}
where $\left(\be^{p_j},\be^{r_j}\right)$ is the representation of the $j$-th post-response pair.

We compute the relevance vector $\bs$ considering two assumptions: (1) topic relatedness can be context-level and word-level; (2) word usage in a post is ambiguous by nature, which biases the topic relatedness. Take the post "Do you like MAC?" as an example. At the context level, it relates to the topic ``personal preference''. At the word level, it relates to the topic ``MAC''. Meanwhile, the word ``MAC'' is ambiguous. For the user in Table~\ref{tab:user}, it has a great chance to be referred to as Apple's "Macintosh" as we can find a ``MAC''-related post ``PC or MAC for college students'' in the dialogue history. But for a beauty blogger, it might refer to cosmetics ``MAC''. 

In view of the first assumption, we decompose the relevance vector $\bs$ into the word-level relevance vector $\bs_1$ and the context-level relevance vector $\bs_2$. 

At word level, for $t$ historical posts with a maximum length of $L$, we obtain the word-level matching matrix $\bM_{w}\in\mathbb{R}^{(t+1)\times L \times L}$ by:
\begin{equation}
    \bM_{w}=\bW^\top_3 \tanh\left(\tilde{\bE}^{P}\cdot \bW_2 \cdot\tilde{\be}^{q\top}\right),
\end{equation}
where $\bW_2 \in \mathbb{R}^{d \times d \times (t+1)}$ and $\bW_3 \in \mathbb{R}^{(t+1) \times 1}$ are parameters. $\tilde{\bE}^P=\{\tilde{\be}^{p_1},\cdots,\tilde{\be}^{p_t},\tilde{\be}^q\}$ is the contextual representations for the current post and historical posts, which is obtained via a single attentive module:  $\tilde{\be}^{(\cdot)} = f_{\text{att}}(\be^{(\cdot)},\be^{(\cdot)},\be^{(\cdot)})$ and $\be^{(\cdot)}$ is initialized by looking up a word embedding table.
We then conduct max-pooling over word-level matching matrix to obtain the most important matching features, which are then linearly mapped into the word-level relevance vector $\bs_1$ using the softmax function: 
\begin{equation}
    \bs_1 = {\rm softmax}\left({\rm MLP}\left(\left[\max_{dim=2}\bM_{w};\max_{dim=3}\bM_{w}\right]\right)\right),
\end{equation}
where $[;]$ is the concatenation operation. 

At context level, we obtain context-level relevance vector $\bs_2$ by:
\begin{equation}
    \bs_2=\frac{\bU^P\cdot \bu^{q}}{\norm{\bU^P}_2\norm{\bu^q}}_2,
\end{equation}
where $\bu^q = \mathop{{\rm mean}}\limits_{dim=1}\tilde{\be}^q$ and $\bU^P = \mathop{{\rm mean}}\limits_{dim=2}\tilde{\bE}^P$ are sentence representation obtained by mean pooling over the word dimension.
We combine word-level and context-level relevance vector by:
\begin{equation}
\label{eq:score}
    \bs=\alpha\cdot \bs_1 +(1-\alpha)\cdot \bs_2,
\end{equation}
where $\alpha$ is a trainable parameter and is initialized by 0.5.

For now, the relevance vector $\bs$ is obtained by using the current query $q$ as a key to attend to historical posts. To tackle the bias discussed in the second assumption, we then design a multi-hop method that alleviates the word ambiguity by stretching the semantic richness of the current query. 
\begin{figure}[t!]
    \centering
    \includegraphics[width=0.8\linewidth]{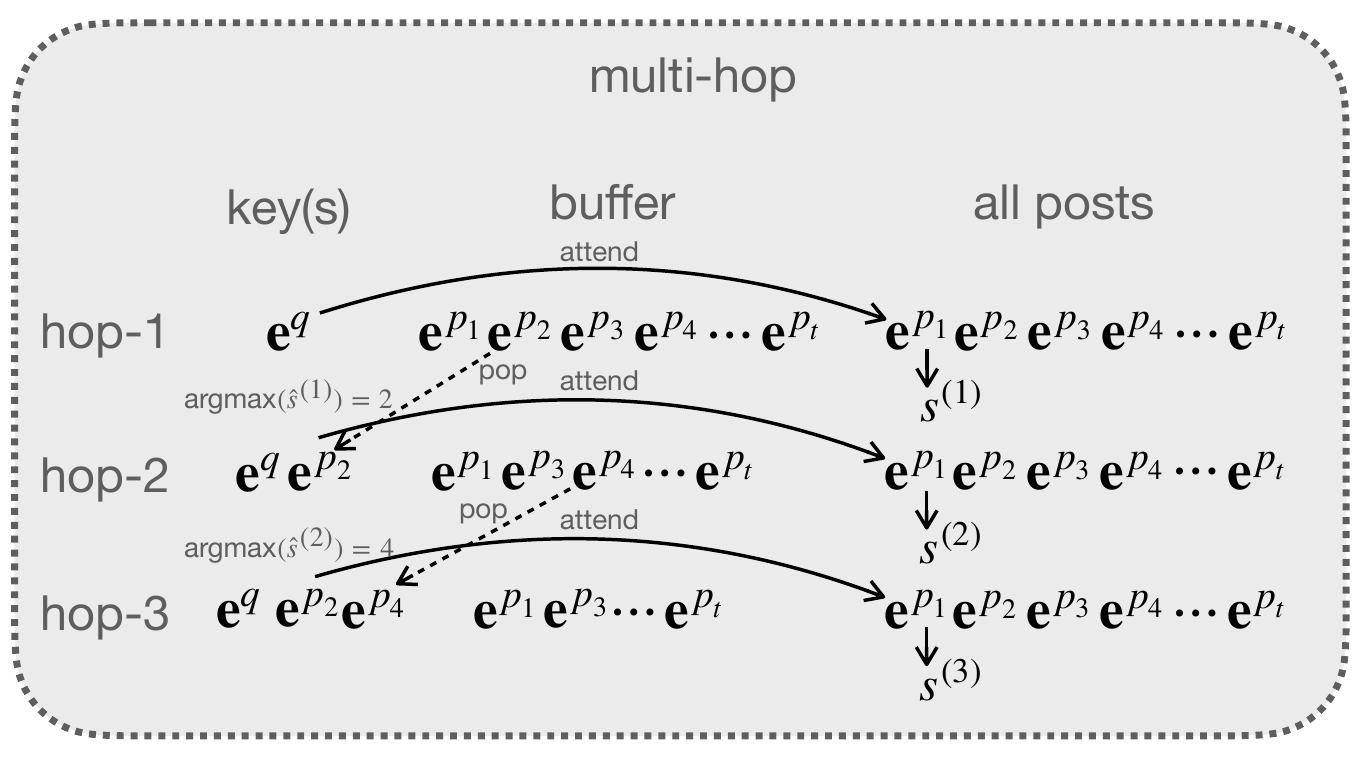}
    \caption{Multi-hop method, we illustrate hop-3. }
    \label{fig:hop}
\end{figure}
Figure \ref{fig:hop} demonstrates the multi-hop method. Specifically, we store the historical posts in a buffer. At each hop, the most related historical post $p_j$ will be popped out as a new key, where $j={\arg\max}(\hat{s})$. $\hat{s}$ is the relevance scores of posts in the buffer. For the post ``Do you like MAC?'' in Table \ref{tab:user}, at hop-1, the key is itself. After attending to historical posts, we expect the historical post ``PC or MAC for college students'' can be popped out as a new key. In this way, the word ``MAC'' is linked to ``PC'' and the ambiguity of the word ``MAC'' can be alleviated. 

We denote all popped keys as $\bE^P_s$. It is a subset of all historical posts $\bE^P=\{\be^{p_1},\cdots,\be^{p_t}\}$. At hop-1, $\bE^P_s=\phi$ and the relevance score is computed by Eq. (\ref{eq:score}) with the current query $\be^q$. Afterwards, we update the current query $\be^q$ by:
\begin{align}
    \be^q= \mathop{{\rm mean}}\limits_{dim=2} f_{\text{stack}}(\be^q\cup \bE^P_{s}).
\end{align}
We then obtain a new relevance score $\bs$ via Eq. \ref{eq:score} using the updated representation $\be^q$. We denote the relevance score of hop-$i$ as $\bs^i$.
After $k$ hops, we have $\bS=\{\bs^1,\cdots,\bs^k\}$. We then linearly map these scores into the final reweighting scores: $\bar{\bs} = \bS \cdot \beta$ where $\beta \in \mathbb{R}^{k\times 1}$ and $\bar{\bs}\in \mathbb{R}^{t+1}$.
Thus, we can rewrite the Eq. (\ref{eq:reweight}) to:
\begin{equation}
    \bD=\bar{\bs}\cdot \bH.
\end{equation}

To thoroughly measure the relevance between the response candidate $r$ and the post-aware user profile $\bD$, we construct three matching matrices: 
\begin{align}
    \bM_1^D&=\left[\frac{\be^{r}\bA_1\bD^{\top}}{\sqrt{d}}; \frac{\be^{r}\cdot \bD^{\top}}{\norm{\be^{r}}_2\norm{\bD}_2}\right], \\
    \bM_2^D&=\left[\frac{\tilde{\be}^{r}\bA_2\tilde{\bD}^{\top}}{\sqrt{d}}; \frac{\tilde{\be}^{r}\cdot \tilde{\bD}^{\top}}{\norm{\tilde{\be}^{r}}_2\norm{\tilde{\bD}}_2}\right], \\
    \bM_3^D&=\left[\frac{\hat{\be}^{r}\bA_3\hat{\bD}^{\top}}{\sqrt{d}}; \frac{\hat{\be}^{r}\cdot \hat{\bD}^{\top}}{\norm{\hat{\be}^{r}}_2\norm{\hat{\bD}}_2}\right], 
\end{align}
where $\bA_*\in\mathbb{R}^{d \times d}$ and $d$ is the embedding size. $\tilde{\be}^{r}$ and $\tilde{\bD}$ are the contextual embeddings obtained via a single-layer attentive module: 
\begin{align}
    \tilde{\be}^{r}&=f_{\text{att}}(\be^{r}, \be^{r}, \be^{r}), \quad \tilde{\bD}=f_{\text{att}}(\bD, \bD, \bD).
\end{align}
Meanwhile, $\hat{\be}^{r}$ and $\hat{\bD}$ are obtained via cross-attention:
 \begin{align}
      \hat{\be}^{r}&=f_{\text{att}}(\tilde{\be}^{r}, \tilde{\bD}, \tilde{\bD}), \quad 
      \hat{\bD}=f_{\text{att}}(\tilde{\bD}, \tilde{\be}^{r}, \tilde{\be}^{r}).
 \end{align}
Thereafter, the three matching matrices are concatenated together:
\begin{align}
    \bM_p=[\bM_1^D;\bM_2^D;\bM_3^D].
\end{align}

Same to the Personalized Style Matching, we use 2D CNN with max-pooling to extract high-level matching features. As the dialogue history is sorted by time, we utilize a single-layer GRU to capture the temporal signal of post-response pairs in the dialogue history.We use the GRU's final state as the Post-Aware Personalized Preference Matching feature $g^{\textbf{P}}\left(q, r, H\right)$. 


\subsection{Module Fusion}
\label{fus}
In Section~\ref{sec:per} and Section~\ref{pro}, we obtain two matching features: (1) personalized style matching feature $g^{\textbf{S}}(q, r, H)$, which measures the personalized style consistency of the response candidates; and (2) post-aware personalized preference matching feature $g^{\textbf{P}}(q, r, H)$, which measures the relevance of a response candidate and the user's personalized preferences.
We concatenate the two matching features to get the final matching vector. We then use an MLP with a sigmoid activation function to compute the final matching score:
\begin{align}
    \mathcal{G} = g\left(q,r,\mathcal{U}\right) = \sigma\left( {\rm MLP}\left(\left[g^{\textbf{S}}\left(q, r, H\right);g^{\textbf{P}}\left(q, r, H\right)\right]\right)\right).
\end{align}


We use cross-entropy loss to train the model:
\begin{align}
    \mathcal{L}(\theta)= -\frac{1}{|D|}\sum\limits_{D}[y\log (\mathcal{G})+(1-y)\log(1-\mathcal{G})].
\end{align}

\section{Experiments}

\subsection{Dataset and Evaluation}

\subsubsection{Datasets}
There are many datasets to evaluate retrieval-based dialogue models~\cite{Lowe2015,Wu2017,Zhang2018, Zhang2018a}. However, none of them contains user identifications. In this paper, we expect to learn implicit user profiles from the user's dialogue history. Thus, we need datasets that contain users' identifications. We use two public datasets crawled from two social networking sites: Weibo and Reddit. 
The Weibo dataset is a subset of the PChatbotW dataset released by~\cite{qian2021pchatbot}. The PChatbotW dataset contains one-year Weibo logs from Sept. 10, 2018. Weibo is a popular social network in China. A Weibo user can post short messages publicly visible, and other users can make responses to the post. In the dataset, all posts and responses have user IDs. 
The Reddit dataset is extracted from the online forum Reddit from Dec. 1, 2015 to Oct. 30, 2018~\cite{DBLP:conf/acl/ZhangSGCBGGLD20}. In the dataset, the discussions can be expanded as tree-structured reply chains where each parent node can be considered as a post to its child nodes. Thus, we generate post-responses pairs by traversing the tree structure. 

\subsubsection{Dataset Construction}
\label{example}
After aggregating users' dialogue history, we filter users with less than fifteen history dialogues to guarantee enough personalized information. Besides, we limit the number of words to 50 for each utterance in the user's dialogue history. For each user, we sort the dialogue history by time and use the latest post as the current query. Following previous works on constructing retrieval-based dialogue datasets~\cite{Lowe2015, Wu2017, Zhang2018a, Zhang2018}, we create a list of ten response candidates for the current query. 

The response candidates can be divided into three groups: (1) we use the user's response under the current query as the ground-truth (personalized response); (2) we select other users' responses under the current query (non-personalized response) as part of the response candidates; (3) following~\cite{Wu2017}, we retrieve response candidates via a retrieval engine (relevant response) and filter out responses issued by the current user as part of the response candidates. Notably, both personalized responses and non-personalized responses can be considered as \textit{proper responses} to the current query.

Compared to previous candidates sampling strategies, such as random sampling~\cite{Lowe2015, Zhang2018a, Zhang2018} or only retrieval~\cite{Wu2017}, our response sampling method is more advantageous, especially regarding personalized chatbot, because:
First, in a practical scenario, the retrieval-based chatbots usually retrieve response candidates relevant to the query via a retrieval engine and then select a proper response from the candidates. Thus, instead of justifying the differences between irrelevant candidates and proper responses, learning to select proper responses from a list of relevant responses is more useful and challenging. Furthermore, we expect that the dialogue chatbot has a consistent personality. Thus, it should also be able to recognize the personalized response from a list of proper responses, including personalized responses and non-personalized ones. The statistical information of the two datasets is shown in Table~\ref{tab:dataset}.

To train baseline models that require external explicit user profile such as {DIM}~\cite{gu2019dually} and {RSM-DCK}~\cite{10.1145/3340531.3411967}, following~\cite{zhong2020towards,DBLP:conf/emnlp/MazareHRB18}, we also collect explicit persona sentences using heuristic methods. Note that these explicit persona sentences are not used in our model.


\begin{table}[t]
\centering
\small
\caption{The statistics of the two datasets. }
\begin{tabular}{lrr}
\toprule
& \textbf{Weibo} & \textbf{Reddit} \\ 
\midrule
Number of users & 420,000 & 280,642 \\
Avg. history length & 32.3 & 85.4\\
Avg. length of post & 24.9 & 10.5\\
Avg. length of response & 10.1& 12.4\\
Avg. number of non-pers. candidates &3.9 & 3.2\\
Avg. number of relevant candidates & 5.1 & 5.8\\
Number of response candidates & 10 & 10\\
Number of training samples & 3,000,000 & 2,000,000 \\
Number of validation samples & 600,000 & 403,210 \\
Number of testing samples & 600,000 & 403,210 \\
\bottomrule
\end{tabular}
\label{tab:dataset}
\end{table}



\subsubsection{Evaluation Metrics}
\label{metrics}
As discussed in Section \ref{example}, the negative response candidates are sampled from the retrieval engine and from other users' responses under the same post (non-personalized response). Thus, we consider three types of metrics to evaluate the models' performances. First, we use $\mathbf{R_n@k}$ (recall at position $k$ in $n$ candidates) and \textbf{MRR} (Mean Reciprocal Rank) to measure the model's ability to select a personalized response from all candidates. Second, we use \textbf{nDCG} (normalized Discounted Cumulative Gain) to measure the model's ability to select a proper response from all candidates. Last, we introduce a new metric $\mathbf{R_p@k}$ to measure the model's ability to select a personalized response from all proper candidates. We introduce the \textbf{nDCG} and $\mathbf{R_p@k}$ in the following:
    
\textbf{nDCG}: the metric assigns scores to both personalized and non-personalized responses. The metric matters in the situation that no personalized response in the list of candidates. In such a scenario, a non-personalized response is better than an irrelevant response. We use \textbf{nDCG@5} and set the relevance score as 2 and 1 for the personalized and non-personalized responses, respectively.

$\mathbf{R_p@k}$: for a personalized chatbot, personalized responses are more useful than non-personalized responses. Thus, we introduce recall at position $k$ in the \textbf{p}roper candidates to measure the model's ability to distinguish the personalized responses from other proper candidates. Taking $\mathbf{R_p@1}$ as an example, when the personalized response ranks 1st out of all proper responses, we assign 1 score to the sample, otherwise 0.

\begin{table*}[t]
\centering
\small
\caption{Evaluation results of all models on both Weibo and Reddit corpus. ``$\dagger$'' and ``$\ddagger$'' denote the result is significantly worse than our method in t-test with $p<0.01$ and $p<0.05$ level respectively. Models with ``$\diamondsuit$'' are implemented with the provided source code, and models with ``$\heartsuit$'' are implemented by ourselves. The best results are in bold.}
\begin{tabular}{lllllllllllll} 
\toprule
  & \multicolumn{6}{c}{Weibo Corpus} &\multicolumn{6}{c}{Reddit Corpus} \\
  \cmidrule(lr){2-7} \cmidrule(lr){8-13}
  & $\mathbf{R_{10}@1}$ &  $\mathbf{R_{10}@2}$ & $\mathbf{R_{10}@5}$ & \textbf{MRR} & \textbf{nDCG} &$\mathbf{R_{p}@1}$ & $\mathbf{R_{10}@1}$ & $\mathbf{R_{10}@2}$ & $\mathbf{R_{10}@5}$ & \textbf{MRR} & \textbf{nDCG} &$\mathbf{R_{p}@1}$ \\  
  \midrule
  (1) {ARC-I}$^\diamondsuit$ & $0.173^\dagger$ & $0.316^\dagger$  &$0.643^\dagger$  &$0.378^\dagger$ &$0.600^\dagger$ &$0.243^\dagger$ & $0.459^\dagger$& $0.597^\dagger$  & $0.848^\dagger$&$0.617^\dagger$ &$0.736^\dagger$ & $0.529^\dagger$    \\
  (1) {ARC-II}$^\diamondsuit$ & $0.178^\dagger$& $0.318^\dagger$ &$0.655^\dagger$ &$0.383^\dagger$  &$0.622^\dagger$ & $0.242^\dagger$  &$0.450^\dagger$ &$0.594^\dagger$ & $0.854^\dagger$&  $0.612^\dagger$ &$0.731^\dagger$ & $0.514^\dagger$   \\
  (1) {KNRM}$^\heartsuit$ & $0.271^\dagger$ & $0.476^\dagger$ & $0.899^\dagger$ & $0.502^\dagger$& $0.856^\dagger$ &$0.277^\dagger$ &$0.558^\dagger$ &$0.721^\dagger$ & $0.905^\dagger$ &$0.698^\dagger$ & $0.847^\dagger$& $0.573^\dagger$   \\
  (1) {Conv-KNRM}$^\heartsuit$   & $0.323^\dagger$ & $0.520^\dagger$ & $0.893^\dagger$ & $0.538^\dagger$ & $0.818^\dagger$ &$0.334^\dagger$&  $0.576^\dagger$&$0.711^\dagger$ &$0.917^\dagger$& $0.712^\dagger$ & $0.797^\dagger$ & $0.616^\dagger$   \\ 
  \midrule
  (2) {SMN}$^\diamondsuit$ & $0.328^\dagger$& $0.525^\dagger$ &$0.894^\dagger$&$0.541^\dagger$  &$0.837^\dagger$ & $0.342^\dagger$ & $0.433^\dagger$ & $0.617^\dagger$ & $0.909^\dagger$ & $0.618^\dagger$  & $0.785^\dagger$ & $0.475^\dagger$  \\ 
  (2) {DAM}$^\diamondsuit$ &$0.438^\dagger$ &$ 0.644^\dagger$ & 0.966 & $0.635^\dagger$ & 0.881 & $0.442^\dagger$ &  $0.605^\dagger$ &$0.748^\dagger$ &$0.965^\dagger$   &$0.741^\dagger$ & $0.830^\dagger$ & $0.626^\dagger$   \\ 
  (2) {IOI}$^\diamondsuit$ & $0.442^\dagger$ & $0.651^\dagger$ &\textbf{0.969}  & $0.639^\dagger$ &\textbf{0.890} & $0.446^\dagger$  & $0.620^\dagger$ & $0.764^\dagger$ &  $0.974^\dagger$ & $0.753^\dagger$ & $0.857^\dagger$ & $0.636^\dagger$  \\
  (2) {MSN}$^\diamondsuit$ &$0.355^\dagger$ & $0.554^\dagger$ & $0.931^\dagger$ & $0.567^\dagger$ & 0.878 & $0.359^\dagger$  & $0.555^\dagger$ & $0.733^\dagger$ & $0.977^\dagger$  & $0.715^\dagger$& $0.875^\dagger$ &  $0.567^\dagger$  \\
  \midrule
  (3) {DIM}$^\diamondsuit$ & $0.388^\dagger$&$0.557^\dagger$  &$0.835^\dagger$  &$0.571^\dagger$ &$0.621^\dagger$ &$0.460^\ddagger$ &  $0.678^\dagger$ & $0.813^\dagger$ &  $0.979^\ddagger$ & $0.794^\dagger$ & $0.804^\dagger$ & \textbf{0.724}  \\ 
  (3) {DGMN}$^\diamondsuit$  & $0.358^\dagger$&$0.528^\dagger$ &$0.818^\dagger$ & $0.547^\dagger$ & $0.636^\dagger$ & $0.424^\dagger$ & $ 0.539^\dagger$ & $0.703^\dagger$ & $0.956^\dagger$ & $0.697^\dagger$ & $0.838^\dagger$ & $0.563^\dagger$  \\
  (3) {RSM-DCK}$^\diamondsuit$  & $0.428^\dagger$ & $0.627^\dagger$ & $0.947^\dagger$ & $0.623^\dagger$ &$0.858^\dagger$ & $0.438^\dagger$  & $0.615^\dagger$ &  $0.753^\dagger$  &$0.972^\dagger$  & $0.748^\dagger$ & $0.838^\dagger$ & $0.633^\dagger$  \\
  (3) {CSN}$^\diamondsuit$  &$0.387^\dagger$ &$0.560^\dagger$  & $0.842^\dagger$ & $0.572^\dagger$ & $0.654^\dagger$ & $0.445^\dagger$  &$0.681^\dagger$ &$0.807^\dagger$ & $0.976^\dagger$& $0.794^\dagger$  &$0.846^\dagger$  & $0.698^\ddagger$  \\
  \midrule
  (3) {IMPChat} & \textbf{0.460}& \textbf{0.665} & 0.963  & \textbf{0.651}& 0.868& \textbf{0.466}  & \textbf{0.691}&\textbf{0.820} & \textbf{0.982}   &\textbf{0.804} &\textbf{0.877}& 0.706 \\
\bottomrule
\end{tabular}
\label{tab:res}
\end{table*}

\subsection{Baseline Model}
In our task, user's dialogue history contains many single-turn dialogues. Given a query, we can either consider the task as single-turn matching or use the dialogue history as a context. Thus, we consider three types of retrieval-based model as baseline models: 

(1) Single-turn matching models:

\textbf{ARC-I}~\cite{hu2014convolutional}: The model uses a Siamese architecture in which multi-layer 1D CNNs capture multi-grained semantic features of each sentence and then perform matching.
\textbf{ARC-II}~\cite{hu2014convolutional}: The model directly builds on the interaction space between two sentences. It uses 1D CNNs to capture low-level features and then uses 2D CNNs to capture deep matching features.
\textbf{KNRM}~\cite{Xiong_2017}: The model uses a kernel-based neural ranking model to model word-level soft matches for single turn-dialogue.
\textbf{Conv-KNRM}~\cite{10.1145/3159652.3159659}: The model uses a CNN kernel-based ranking model to model n-gram soft matches for single-turn dialogue.
    
(2) Multi-turn dialogue models:

\textbf{SMN}~\cite{Wu2017}: The model matches a response with each utterance in the context on multi-level. It uses CNN to extract matching information and obtains the final matching score using an RNN.
\textbf{DAM}~\cite{zhou-etal-2018-multi}: The model constructs multi-level text segment representations with stacked self-attention and then extracts the matched segment pairs with attention across the context and response. 
\textbf{IOI}~\cite{tao-etal-2019-one}: The model uses a chain of interactive blocks to conduct semantic interaction between response and utterance in the context many times to obtain deep interactive matching information.
\textbf{MSN}~\cite{yuan-etal-2019-multi}: The model conducts context selection and filtered irrelevant context. It lets the response candidates interact with each utterance in the context to get multiple matching features. 


(3) Persona-based dialogue models:

\textbf{DIM}~\cite{gu2019dually}: The model uses BiLSTM to encode response candidates, user profile, and context, and lets the user profile, context interact with response candidates, respectively, via cross-matching. It utilizes another BiLSTM to aggregate the matching features.
\textbf{DGMN}~\cite{DBLP:conf/ijcai/ZhaoTWX0Y19}: The model fuses information in a document and a context into representations of each other. It performs hierarchical interactions between a response and both document and context. And the importance of each part is dynamically determined.
\textbf{RSM-DCK}~\cite{10.1145/3340531.3411967}: The model pre-selects document and context. It performs matching between response-context and response-document. It uses Bi-LSTM to aggregate matching features.
\textbf{CSN}~\cite{ZhuNZDD21}: The model designs a content selection network to explicitly select relevant contents, and filter out the irrelevant parts. It performs context-response matching and document-response matching and aggregates matching features using CNN with LSTM.

\subsection{Implementation Details}
In our model, we use Word2Vec \cite{mikolov2013efficient} to initialize the word embedding, which has a size of 200. The Word2Vec is trained on the dataset. In the experiments, all baseline models apply the word embedding. We limit the max sequence length to 50. For the Personalized Style Matching module, we set the number of attentive modules to 3. For the Post-Aware Personalized Preference Matching module, we set the number of hops to 2 and set the hidden size of GRU to 300. We use three-layer CNN with $16 [3,3]$, $32 [3,3]$, $64 [3,3]$ filters, respectively. The 1st and 2nd layer use $[2,2]$ stride and the max-pooling size is $[2,2]$ with $[2,2]$ stride. The 3rd layer uses $[3,3]$ stride, and the max-pooling size is $[3,3]$ with $[3,3]$ stride. We optimize the model using the Adam method with a learning rate set as 5e-4, which is decayed during training. We tune IMPChat and all baseline models on the validation set and evaluate on the test set. For both datasets, we set batch size as 128 and train the model for 10 epochs on two Tesla V100 16G GPUs. Our model is implemented by Pytorch~\cite{NEURIPS2019_9015}, and the code will be released based upon the acceptance of the paper.

\subsection{Experimental Results}
Table~\ref{tab:res} shows the results. Compared to previous state-of-the-art models, we find that \textbf{our model IMPChat achieves the best performance across most of the metrics on the two datasets}. As mentioned in Section~\ref{metrics}, we use three types of evaluation metrics that reflect the model's abilities from different aspects. \textbf{First}, regarding ${R_{10}@1}$, ${R_{10}@2}$ and {MRR}, our model IMPChat leads to statistically significant improvement than all baseline models on the two datasets (t-test with $p<0.01$). It demonstrates the strong capability of our IMPChat to distinguish the marginal differences between personalized responses and all other responses. \textbf{Second}, regarding {nDCG}, IMPChat outperforms all baselines on Reddit Corpus while IMPChat is worse than several multi-turn models on the Weibo dataset. We also find that most other personalized models perform worse than multi-turn models in terms of {nDCG}. The potential reasons might be: (1) personalized information increases the marginal distance between personalized responses and other responses. Therefore, the marginal distance between non-personalized (but relevant) responses and other negative responses decreases; (2) more personalized information also brings more noise that undermines matching (also discussed in section \ref{length}). \textbf{Last}, regarding ${R_{p}@1}$, we find that most personalized models perform better than multi-turn models, reflecting that personalized information enhances the models' capability to distinguish personalized responses from non-personalized responses. Our IMPChat outperforms all baseline models on Weibo Corpus and baseline models excluding DIM on Reddit Corpus, showing the effectiveness of building implicit user profiles from the dialogue history.

\subsection{Discussion}
\subsubsection{Ablation Study}
We respectively remove (1) the multi-hop method, (2) the Personalized Style Matching module, and (3) the Post-Aware Personalized Preference Matching module of IMPChat to investigate their effectiveness. Table~\ref{tab:abl} shows the results. We can conclude that removing any part of IMPChat will lead to a performance drop. Notably, performance degrades a large margin when removing the Post-Aware Personalized Preference Matching module. This is because, without the module, IMPChat only accesses historical responses so that the conditional matching signals among post-response pairs diminish. Besides, we find that without either multi-hop or Personalized Style Matching module, IMPChat still outperforms all baseline models regarding ${R_{10}@1}$, ${R_{10}@2}$, {MRR} and ${R_{p}@1}$ which verifies that either of the two parts indeed 
brings orthogonal information that the rest parts fail to capture.


\begin{table}[t!]
\centering
\small
\caption{Ablation results on the Weibo dataset.}
\setlength{\tabcolsep}{1.4mm}{
\begin{tabular}{lcccccc} 
\toprule
& $\mathbf{R_{10}@1}$ &  $\mathbf{R_{10}@2}$ & $\mathbf{R_{10}@5}$ & \textbf{MRR} & \textbf{nDCG} &$\mathbf{R_{p}@1}$\\ 
\midrule
  {IMPChat}  &\textbf{0.460} & \textbf{0.665} & 0.963 & \textbf{0.651} & 0.868 & \textbf{0.466} \\ 
  \quad{\textit{w/o} mutil-hop}  &0.451 &0.662 &\textbf{0.966} &0.646 &\textbf{0.878} &0.455  \\ 
  \quad{\textit{w/o} style}  & 0.443 & 0.654 & 0.962 & 0.639 &0.871  &0.447  \\ 
  \quad{\textit{w/o} preference}  & 0.383&0.572 &0.939 &0.585 &0.867 &0.391  \\ 
\bottomrule
\end{tabular}
}
\label{tab:abl}
\end{table}

\subsubsection{How to Choose the History Length?}
\label{length}
\begin{figure}[t!]
    \begin{subfigure}[b]{0.49\linewidth}
        \centering
        \includegraphics[width=\textwidth]{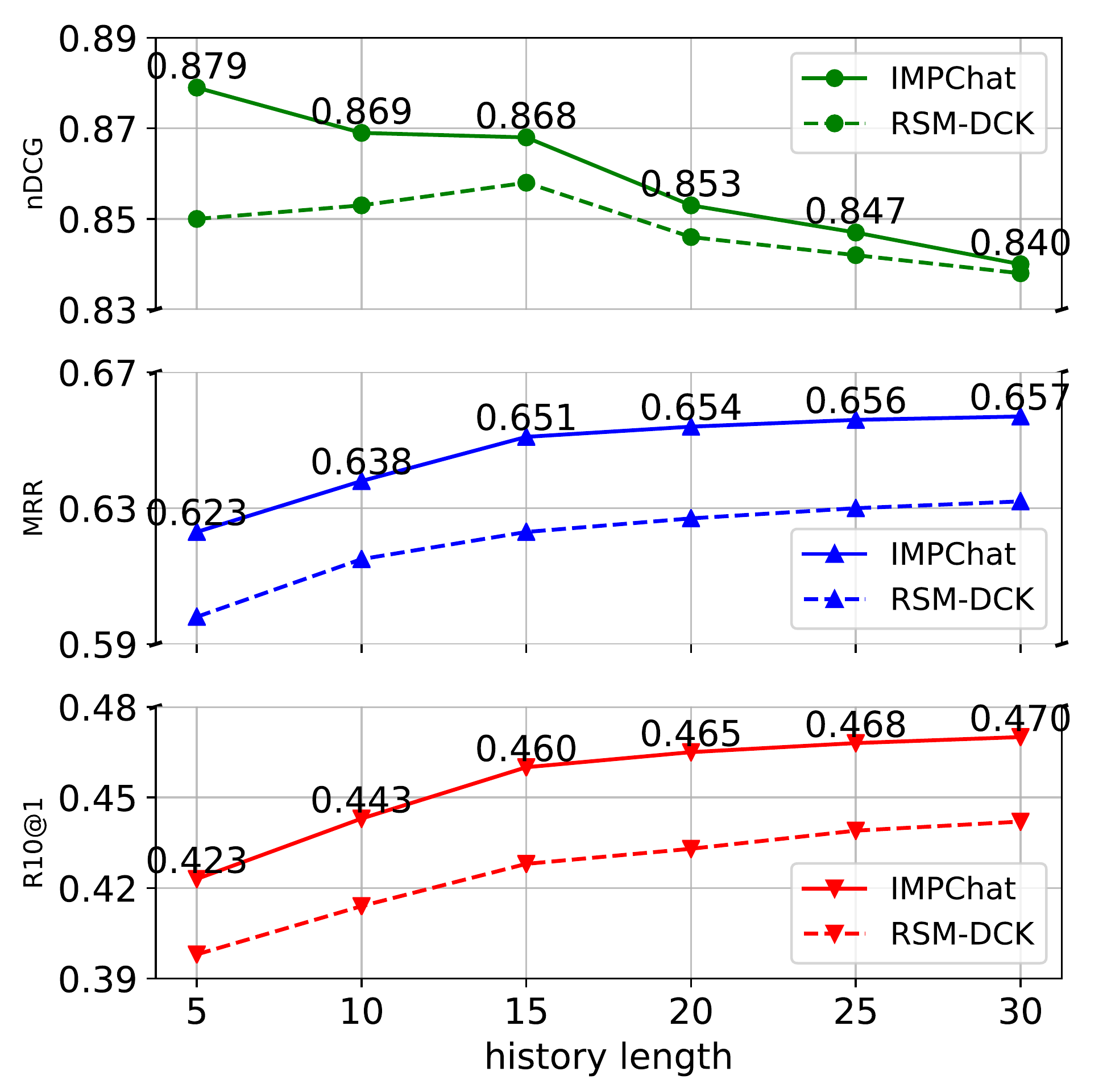}
        \caption{Effect of history length.}
        \label{fig:perfo}
    \end{subfigure}
    \hfill
    \begin{subfigure}[b]{0.49\linewidth}
        \centering
        \includegraphics[width=\textwidth]{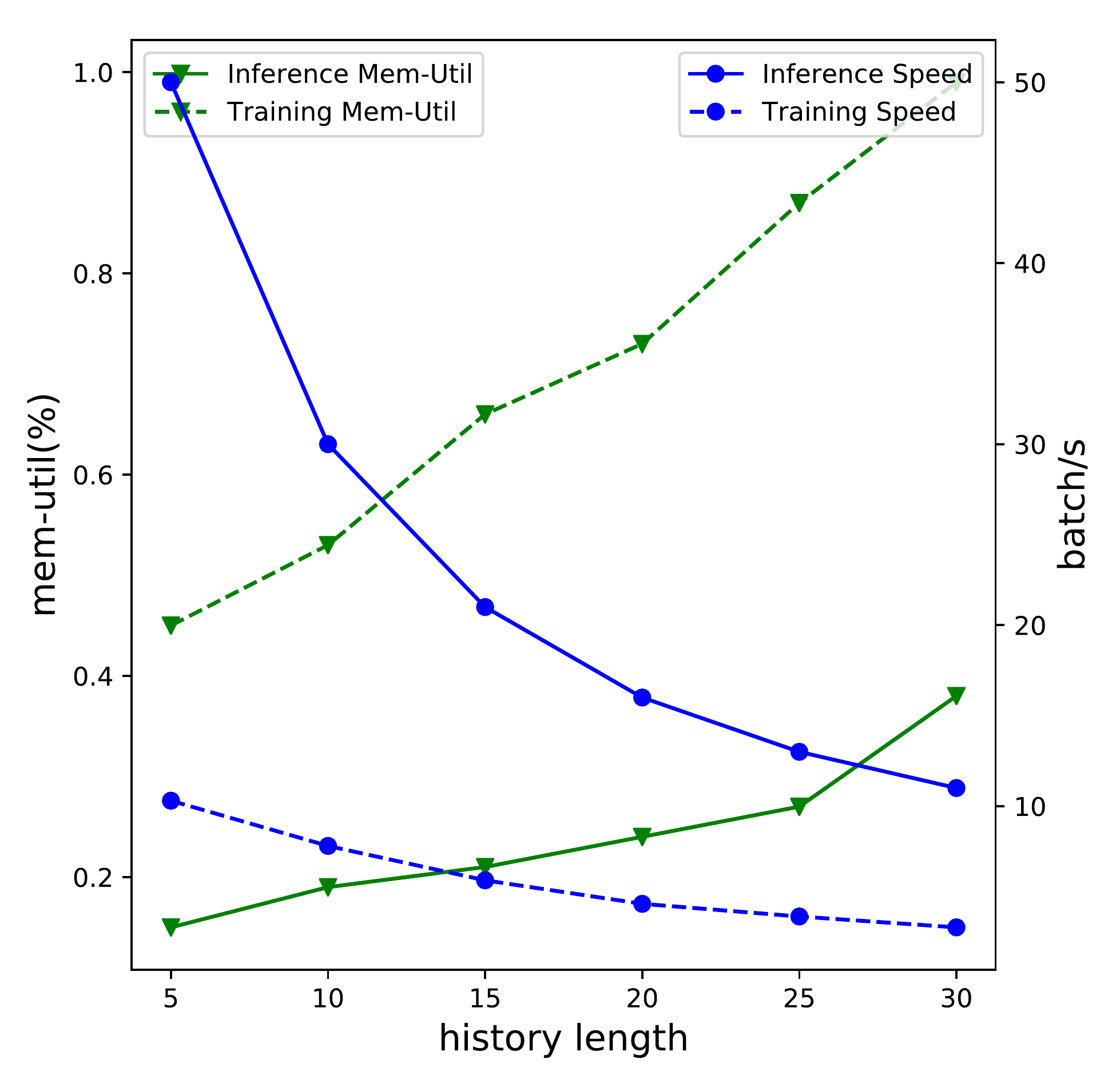}
        \caption{Resource usage and speed.}
        \label{fig:res}
    \end{subfigure}
    \caption{The left figure shows the performance variance on different history lengths. The right figure shows the required GPU memory and speed for training and inference.}
    \label{fig:length}
\end{figure}

We aim to learn an implicit user profile from the user's dialogue history. Thus, how to choose the length of dialogue history is a crucial problem. We investigate the problem from two aspects: how the history dialogue length affects the model's performance; and the difficulty to train the model or make inferences with more dialogue history. Figure~\ref{fig:length} shows the influence of the history length, from which we can find: (1) as aforementioned, ${R_{10}@1}$ and {MRR} measure the model's capability to retrieve a personalized response, while {nDCG} also considers non-personalized responses. Thus, with more dialogue history, the model's capability of finding the personalized response steadily increases since the longer history brings more personalized information. However, the model's capability to retrieve non-personalized responses decreases. The potential reason for the decline is that long history brings more noise into the implicit profile, which ambiguates the decision boundaries between non-personalized responses and other negative ones; (2) Figure~\ref{fig:res} shows that the required computing resources increase with longer dialogue history\footnote{We set the batch size to 32, and conduct experiments on a single Tesla V100 16G GPU.}. In a practical scenario, there is a trade-off regarding how to choose the length of dialogue history. First, training and inference speeds matter in practical use, especially for devices with limited computing resources such as a mobile phone. Second, not always personalized responses can be retrieved before response selection. When no response matches the user's personality, a relevant response is better than a wrong one. We also conduct experiments on RSM-DCK, which has a similar number of parameters to IMPChat. The results also verify our findings.

\section{Conclusion}
In this paper, we propose building implicit user profiles to endow personality to a retrieval-based chatbot. To achieve such a goal, we build a model IMPChat to learn the implicit user profile from two aspects. First, it models the user's personalized language style using the user's historical responses. It then models the user's personalized preferences from a post-aware user profile which contains post-response pairs that are topically related to the current post. Extensive experimental results on two large datasets show that our method outperforms all previous state-of-the-art models, verifying our model's effectiveness for the personalized retrieval-based chatbot. 

\section*{Acknowledgement}
Zhicheng Dou is the corresponding author. This work was supported by National Natural Science Foundation of China No. 61872370 and No. 61832017,  Beijing Outstanding Young Scientist Program NO. BJJWZYJH012019100020098, Shandong Provincial Natural Science Foundation under Grant ZR2019ZD06, and Intelligent Social Governance Platform, Major Innovation \& Planning Interdisciplinary Platform for the "Double-First Class" Initiative, Renmin University of China. I also wish to acknowledge the support provided and contribution made by Public Policy and Decision-making Research Lab of Renmin University of China.

\balance
\clearpage
\bibliographystyle{ACM-Reference-Format}
\bibliography{sample-base}


\end{document}